# Large magnetoresistance anomalies in $Dy_7Rh_3$

**Kausik Sengupta, S. Rayaprol, and E.V. Sampathkumaran**
Tata Institute of Fundamental Research, Homi Bhabha Road, Mumbai 400 005, India.

### *Abstract*

The compound $Dy_7Rh_3$ ordering antiferromagnetically below ($T_N=$) 59 K has been known to exhibit a temperature (T) dependent electrical resistivity ($\rho$) behavior in the paramagnetic state unusual for intermetallic compounds in the sense that there is a broad peak in $\rho$(T) in the paramagnetic state (around 130 K) as though there is a semi-conductor to metal transition. In addition, there is an upturn below $T_N$ due to magnetic super-zone gap effects. Here we report the influence of external magnetic field (H) on the $\rho$(T) behavior of this compound below 300 K. The rise of $\rho$(T) found below $T_N$ could be suppressed at very high fields (>> 60 kOe), thus resulting in a very large magnetoresistance (MR) in the magnetically ordered state. The most notable finding is that the magnitude of MR is large for moderate applications of H (say 80 kOe) in a temperature range far above $T_N$ as well, which is untypical of intermetallic compounds. Thus, this compound is characterized by large MR anomalies in the entire T range of investigation.

During the last decade, there have been considerable efforts in identifying materials exhibiting large magnetoresistance (MR), particularly near room temperature, considering its application potential. Such reports are however scarce among intermetallics due to the fact that the net magnitude of MR caused by the influence of external field (H) on the cyclotron motion of the conduction electrons as well as on the paramagnetic moments is usually negligible in the vicinity of room temperature. In this article, we report the influence of H on the temperature (T) dependent electrical resistivity ($\rho$) behavior of an intermetallic compound, $Dy_7Rh_3$, which is found to exhibit interesting magnetoresistive response in the entire temperature of investigation. It may be remarked that the compounds of the type, $R_7Rh_3$ (R= Rare-earths), crystallizing in the $Th_7Fe_3$-type hexagonal structure ($P6_3mc$), are of an unusual type among the rare-earth intermetallics in the sense that the light R members behave like metals, whereas in heavier members there are temperature regions in which the T-coefficient of $\rho$ is negative not only in the magnetically ordered state but also in the paramagnetic state [1-3]. Such features are believed to arise from the existence of a band-gap, which is sensitive to lanthanide contraction effects in this series.

Polycrystalline sample of $Dy_7Rh_3$ was synthesized by arc-melting high purity constituent elements in argon atmosphere. The molten ingot was annealed at $600^o$ C for 2 days in vacuum and characterized by x-ray diffraction. Powder x-ray diffraction pattern established that the sample is single phase and the lattice parameters are in close agreement with the reported ones [3]. The dc $\rho$ measurements were performed in the T interval 1.5-300 K by a four-probe method in the presence of various H employing the Physical Property Measurements System (Quantum Design). In addition, the measurements were performed as a function of H at selected temperatures.

In Fig. 1, temperature dependence of $\rho(T)/\rho(300 K)$ is plotted. The value of $\rho$ at 300 K is typically of the order of 300 $\mu\Omega$cm. It is obvious that, for H= 0, the sign of $d\rho/dT$ is clearly negative above 150 K due to a semimetallic band gap of about 9 meV [3]; around 130 K there is a broad peak followed by a gradual fall as the T is further lowered in the paramagnetic state as though there is a semimetallic-metallic continuous transition. There is a sudden upturn in $\rho$ near 50 K after the onset of antiferromagnetic ordering due to the formation of magnetic super-zone gap. While these findings are in broad agreement with those reported in the literature on polycrystals [3], a drop seen below 34 K in Ref. 3 is absent in our samples in agreement with the features exhibited by the single crystalline form [4]. We find that the features are qualitatively unaffected by the application of H in the paramagnetic state, though the maximum appears to shift to a higher T range with increasing field, say to 200 K at 140 kOe. These results imply that the paramagnetic transport behavior is robust to the application of H. As the sample enters magnetically ordered state, the $\rho(T)$ behavior qualitatively remains unaffected till H= 80 kOe, while for H= 140 kOe, the upturn vanishes as though the magnetic gap can be suppressed by such large fields only.

We low look at the trends observed in the absolute values of $\rho$ as H is varied. For this purpose, we have derived the magnitudes of MR (= $[\rho(H)- \rho(0)]/\rho(0)$) as a function of T. The results are shown in figure 2. Clearly, for H= 50 and 80 kOe, the magnitude of MR increases with decreasing T attaining a peak near the magnetic transition (as large as –13 % for H= 80 kOe), while the magnitude keeps growing down to very low temperatures for H= 140 kOe attaining a value of about -45% at 1.8 K. The magnitude remains large (-28%) even at temperatures as high as 50 K for 140 kOe. The large MR with a negative sign in the high-field range originates from the reconstruction of Fermi surface due to the collapse of the magnetic super-zone gap resulting in a recovery of the carrier concentration.

In order to have a closer look at the magnitudes of the MR, we show in figure 3 the values of MR at several temperatures, obtained by measuring $\rho$ as a function of H. In the magnetically ordered state, though the response of $\rho$ to initial applications of H is rather negligible, it is clear that the magnitude becomes large at higher fields as noted above. Another finding we make is that there are steeper changes in MR near 60 and 120 kOe for T= 1.8 and 20 K, and these are found to be hysteretic. We attribute these to the existence of metamagnetic transitions (first-order-like, but presumably broadened by defects), inferred from the isothermal magnetization studies on single crystals in Ref. 4. But, such a transition is absent at 50 K as evidenced by the present data seen in Fig. 3; the shape of MR plot at 50 K is in fact different from those observed at lower temperatures, which implies that the magnetic structure varies as T is lowered from 50 to 20 K.

We now focus on the MR behavior in the paramagnetic state (see, for instance, Fig. 3, bottom). We find that the sign of MR is negative and that MR varies essentially quadratically (see Fig. 4) for moderate applications of H, typical of the influence of H on the paramagnetic moments; for large enough field values, there is a deviation from quadratic behavior possibly due to an influence of H on the Fermi surface (on the band gap). The most interesting point is that the observed magnitudes of MR in the high field region are reasonably large. Thus, even at temperatures as high as 200 K and 150 K, the values are about –3% and –5.5% at 140 kOe. Possibly, the influence of H on the band-gaps at the Fermi surface also plays a role to determine large magnitude of MR. We hope that this observation will motivate further investigations in this regard.

Summarizing, the intermetallic compound $Dy_7Rh_3$ exhibits interesting temperature dependent $\rho$ behavior, the qualitative features of which are quite robust to the presence of moderate magnetic fields (say, 80 kOe). However, the absolute values of MR are quite large not only in the magnetically ordered state (below 50 K), but also in the paramagnetic state for H beyond certain values. The values observed in the range 100 to 200 K are comparable to those in few other intermetallic compounds [5,6], which have been classified as *metallic* giant magnetoresistance systems in this temperature range. While the MR anomaly arises in the paramagnetic state itself [7] in the present compound, for the latter compounds, the anomalies are essentially in the magnetically ordered state due to field-induced changes in the spin orientation. In that sense, the present compound presents a new situation.

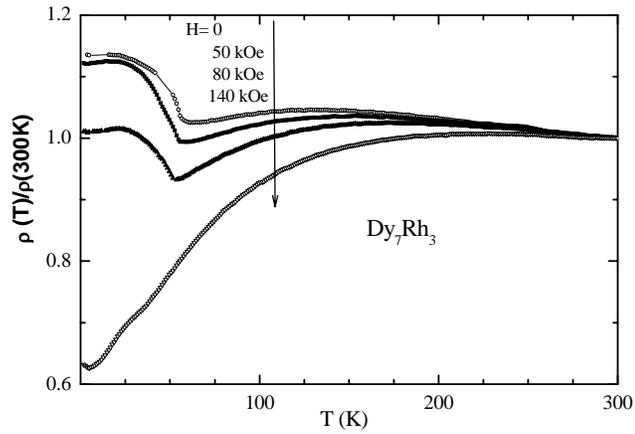

Fig. 1. Electrical resistivity of Dy$_7$Rh$_3$ as a function of temperature measured in the presence of magnetic fields, normalized to respective 300 K values.

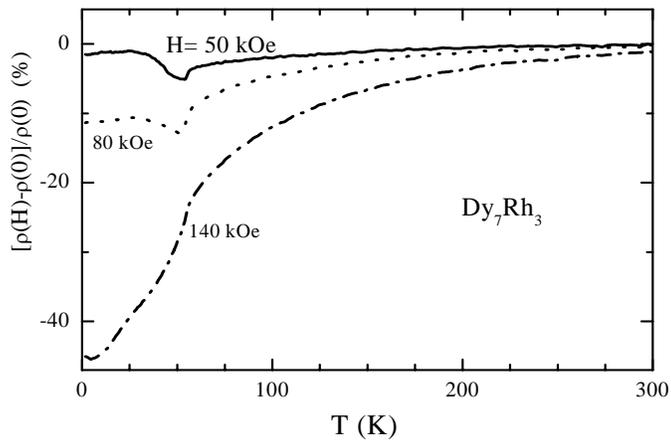

Fig. 2: The magnetoresistance as a function of temperature for Dy$_7$Rh$_3$ derived by taking the resistance data at fixed H values.

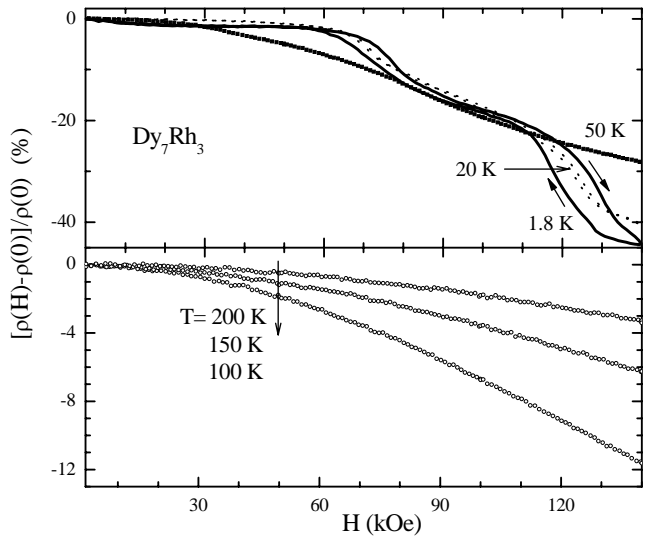

Fig. 3. Magnetoresistance of Dy$_7$Rh$_3$ measured at selected temperatures.

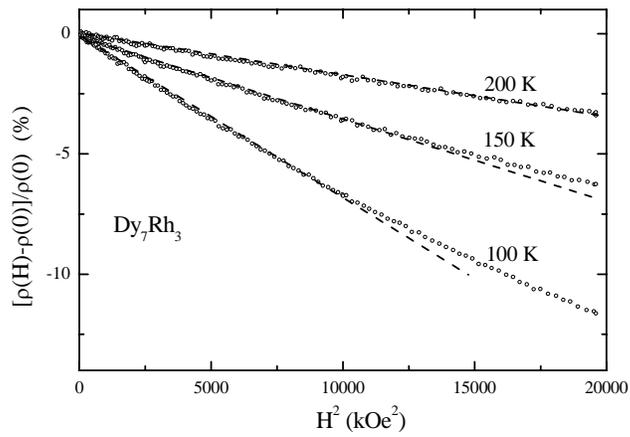

Fig. 4: The plot of magnetoresistance vs square of temperature for Dy$_7$Rh$_3$ at various temperatures in the paramagnetic state. Dotted lines are drawn through the quadratic region.